\documentclass[preprint,aps,prd,floats,floatfix,amsmath,amssymb,amsfonts,nofootinbib,longbibliography,superscriptaddress
]{revtex4-2}

\usepackage{cases,float,physics}
\usepackage{todonotes}
\usepackage{graphicx}
\usepackage{dcolumn}
\usepackage{bm}
\usepackage{hyperref}

\usepackage{amsfonts,amsmath,amssymb}
\usepackage{comment}
\begin{document}

\preprint{APS/123-QED}

\title{{\bf Surface scalar plasmons on curved interfaces: a venue for analogue gravity}}

\author{Matheus E. Pereira}
\email{mepereira@id.uff.br}
 \affiliation{%
  Instituto de Ci\^encias Exatas, Universidade Federal Fluminense,
  \\ 27213-145 Volta Redonda --- RJ, Brasil
}%



\date{\today}

\begin{abstract}
We investigate whether the geometry of a curved conductor--dielectric interface can modify the effective mass and propagation of scalar surface plasmons (SSPs). In the hydrodynamic limit, we describe SSPs as density perturbations in a semiclassical jellium electron gas. Using a thin--layer dimensional reduction procedure, we derive a  massive Klein--Gordon equation for SSPs on the curved interface with a geometric potential. We find that under specific conditions, the geometric potential allows SSP propagation with a vanishing effective mass. We then apply the formalism to a well known analogue black hole configuration, determining the corresponding quasinormal mode frequencies. Our results show that interface curvature provides a geometric handle on plasmon propagation and localization, suggesting new possibilities for plasmon transport and analogue gravity experiments with surface plasmons.
\end{abstract}

\maketitle


\section{Introduction}\label{sec:introducao}

The parallels between the mathematical frameworks of General Relativity and physical systems in fields such as condensed matter physics and optics have been used to advance our understanding of both, revealing two complementary sides of the same analogy. On the one hand, concepts and results from General Relativity can be used to explore new possibilities in laboratory systems, as exemplified by the works of Zel'dovich, Unruh \cite{zeldovich1,zeldovich2,unruh01}, and Hou et al. \cite{hyperlente}. On the other hand, laboratory systems can be employed to investigate the physical realization of these mathematical analogies. An important example of the latter is the experimental observation of Hawking radiation, originally proposed by Unruh \cite{unruh01,unruh-shallow} and subsequently reported by Weinfurtner and colleagues \cite{weinfurtner2011}.

Originally motivated by the possibility of investigating Hawking radiation through analogue experiments \cite{unruh01}, Analogue Gravity has since evolved into a broader research program. Beyond providing experimental evidence for radiation emission from artificial horizons \cite{analogue-gravity-review,leonhardt-original,leonhardt2008,weinfurtner2011, Rubino_2011,Drori2019, fischer-petrov}, it has also opened avenues for laboratory simulations \cite{jacquet2020next} of a variety of gravitational phenomena, including effects of interest in astrophysics \cite{Hartley_gravwaves} and cosmology \cite{barcelo2003analogue,fedichev2004cosmological,lidsey2004cosmic,Weinfurtner2007,prain2010analogue,faraoni2020lagrangian,velten2012power,de2015nonlinear,fifer2019analog,steinhauer2022analogue,niehof2026freezing,faraoni2026laboratory,hyperlente, cosmologia,Chandran_2025,fischer-ralf}. On the other hand, the dynamics of a quasiparticle in condensed matter systems have been explored and applied to the Analogue Gravity area using phonons in Bose--Einstein condensates \cite{fischer-ralf,prain2010analogue}, polariton fluids in microcavities \cite{jacquet-polariton}, magnons \cite{tese-magnon}, Goldstone modes in liquid crystals \cite{liquid-crystal}, and more. 

In this work we will focus on yet another quasiparticle --- the scalar surface plasmon, or SSP. We work in the electrostatic/longitudinal limit, where the plasmon is described by a scalar density perturbation and the magnetic field is neglected; hence we refer to them as scalar surface plasmons. Plasmons in a jellium model, which is a phenomenological, and in this case, semiclassical model of electrons embedded in a continuous, uniform positive background that maintains overall charge neutrality, allowing electron--electron interactions to be investigated without explicitly accounting for the underlying ionic lattice, have been used before in Analogue Gravity \cite{kolomeisky2019natural, kolomeisky2020normal, kolomeisky2021analog, matheus-jellium} in an analogy to Newtonian cosmology. While we also employ a jellium model, we focus on the surface dynamics rather then the bulk evolution, in contrast with these previous works. Introducing a curved interface adds a genuinely geometric ingredient to the problem: an effective geometric potential, first introduced by \cite{dacosta} and later generalized by Ferrari and Cuoghi \cite{Ferrari_2008}, that evidences suggest to play a role in particle localization \cite{valle}. 
%

Our goal is to investigate the following question: can the geometry of a curved conductor--dielectric interface modify the effective mass and propagation of surface plasmons in a way that can be exploited within Analogue Gravity? To address this question, we investigate the dynamics of scalar surface plasmons (SSPs) on curved interfaces, building on the dimensional reduction framework developed by da Costa and subsequent extensions \cite{tese-renato,tese-matheus,longhi,natalia1,natalia2,geometrohydro}. In particular, we apply a similar thin--layer procedure to a semiclassical electron gas described within a hydrodynamic regime and determine the geometric potential that emerges in the resulting effective surface theory. We then examine the consequences of this potential for plasmon propagation and its possible application to a well studied analogue black hole configuration \cite{analogue-gravity-review}, motivated by previous investigations \cite{kerr-de-sitter,kerr-sen,dini,matheus-jellium}.

Following da Costa's approach \cite{dacosta}, we perform the dimensional reduction of the density perturbations in the electron gas and impose a boundary condition that localizes the plasmons near the curved interface. The resulting surface equation is of Klein--Gordon type and contains both a mass--like term and a geometric potential. We find that this geometric potential differs from the one obtained in the original thin--layer formulation of da Costa and, under specific conditions, can compensate the effective mass--like contribution, allowing for plasmon propagation with a vanishing effective mass. We further apply the resulting formalism to an analogue black hole configuration and extend a numerical method introduced by \cite{fiziev2006} to determine the corresponding frequency eigenvalues under quasinormal mode (QNM) boundary conditions.

The paper is organized as follows. Section \ref{sec:fluido} introduces the charged fluid description and, through a multiscale perturbative analysis, derives the wave equation governing the SSPs. Section \ref{sec:da costa} then develops the thin--layer reduction and obtains the corresponding geometric potential, as well as a new effective metric tensor, directly related to the shallow--water results \cite{unruh-shallow} commonly employed in analogue gravity experiments. Its physical interpretation and implications for plasmon localization are discussed in Section \ref{sec:interpretacao}. Section \ref{sec:analogue} applies the formalism to an analogue gravity configuration. The QNM analysis and numerical results are presented in Section \ref{sec:resultados}. Finally, Section \ref{sec:conclusao} summarizes our results and discusses the physical implications of our geometric potential, which is different from da Costa's.

\section{Analogue gravity in a charged fluid}\label{sec:fluido}
In a dispersionless conducting medium, we consider the electron fluid in the hydrodynamic regime with all ions essentially still, i.e., a jellium, whose particles have mass density $r_m = m_e n$ and charge density $-e(n - n_B)$, where $e$ is the electron charge, $m_e$ is the electron mass, $n_B$ is the background electron density per unit volume, and $n$ is the number density of electrons in the fluid, can be described by \cite{maier2007plasmonics, haas2011quantum, mendonca, manfredi2015solid, barcelo2003analogue,schroeder2021introduction}
	\begin{align}
	&\partial_{t} n +\vec{\nabla}\cdot(n \mathbf{v})=0 \label{eq continuidade plasmons},\\
	&m_e n\left(\partial_{t} \mathbf{v}+\left(\mathbf{v} \cdot \vec{\nabla} \right)\mathbf{v}\right)=-\vec{\nabla} p - m_e n \Gamma\mathbf{v} - n e \mathbf{E} - n e \mathbf{E}_\text{ext}, \label{eq de euler plasmons} \\
    &\vec{\nabla} \cdot \mathbf{E}= -\frac{e(n - n_B)}{\varepsilon} ,\label{lei de gauss plasmons}
	\end{align}	
where $\varepsilon$ is the permittivity of the medium and we take the fluid to be collisionless, i.e., $\Gamma = 0$. We assume that gravitational forces are negligible and that $-\vec{\nabla}V_{\mathrm{el}} = \mathbf{E}$. Furthermore, if the flow is irrotational, i.e., $\vec{\nabla} \times \mathbf{v}=0$, the velocity field can be written as $\mathbf{v} = \vec{\nabla}\psi$. Following the usual steps of the perturbation method, we set $\delta$ as an unitless parameter to control the amplitudes of our fields and use expansions $n \approx n_{0} + \delta n_{1} + O(\delta^2), 
            \mathbf{v} \approx \mathbf{v}_{0} + \delta\mathbf{v}_{1} + O(\delta^2)$, and the same for the pressure $p$ and the electric field $\mathbf{E}$. The case of an incompressible background velocity field simplifies the zeroth--order continuity equation to
\begin{equation}\label{velocidade incompressivel}
    \vec{\nabla}\cdot\mathbf{v}_0 = 0.
\end{equation}
Since we also require the flow to be irrotational, there exists a background velocity potential $\psi_0$ such that $\mathbf{v}_0 = \vec{\nabla}\psi_0$ and $\nabla^2 \psi_0 = 0$. After a few mathematical steps, the system becomes
\begin{numcases}
        {} \vec{\nabla}\cdot\mathbf{v}_{0} = 0 , \label{eq continuidade ordem 0 n constante sistema 3} \\
		  \partial_{t} n_{1} + n_0 \vec{\nabla}\cdot \mathbf{v}_{1} + \mathbf{v}_{0} \cdot \vec{\nabla} n_{1} = 0 ,\label{eq continuidade ordem 1 n constante sistema 3}  \\
          \left[\partial_{t}  +  \left(\mathbf{v}_{0}\cdot \vec{\nabla}\right)\right] \mathbf{v}_{0}= -\frac{1}{m_e n_0}  \vec{\nabla} p_{0} -  \frac{e}{m_e}\left( \mathbf{E}_{0} + \mathbf{E}_\text{ext}\right), \label{eq euler ordem 0 n constante sistema 3}  \\
		  \left[\partial_{t} +
       \left(\mathbf{v}_{0} \cdot \vec{\nabla}\right) \right]\mathbf{v}_{1} 
      +  \left(\mathbf{v}_{1} \cdot  \vec{\nabla}\right) \mathbf{v}_{0}
       = - c_s^2 \vec{\nabla}\left(\frac{n_1}{n_0}\right) - \frac{e}{m_e}\mathbf{E}_{1}, \label{eq euler ordem 1 n constante sistema 3} 
\end{numcases}
up to linear order.
These are the fundamental equations that will be used to describe a wave-like evolution of our perturbations. These perturbations we are interested in give rise to longitudinal oscillations in the electron fluid, whose wavelength $\lambda$ satisfies the hierarchy $\lambda_D \ll \lambda \ll H^{-1}$, where $H^{-1}$ is the characteristic length scale over which the background varies, and $\lambda_D$ is the Debye length, defined as
    \begin{equation}
        \lambda_D = \sqrt{\frac{\varepsilon k_B T}{n e^2}}.
    \end{equation}
This wavelength \cite{kittel2018introduction, kittel1963quantum,Stenson_Horn-Stanja_Stoneking_Pedersen_2017} is a function of the density $n$ and the temperature $T$, where $k_B$ is the Boltzmann constant. Additionally, we make use of the Drude model for the electric conduction in a medium. In this model, the 

We introduce the parameter $\sigma=\lambda H\ll 1$, which sugests a new scale of the wavelength variations, independent of $\delta$. Thus, we define a \textit{slow} spatial variable $\boldsymbol{\xi}$ as $\boldsymbol{\xi} = \sigma\mathbf{x}$ and a \textit{slow} time variable $\tau$ as $\tau = \sigma t$, to be used in a multiple scale analysis \cite{valle,chini}. By the chain rule, we obtain a general transformation
\begin{equation}
    \vec{\nabla} \to \vec{\nabla}_\mathbf{x} + \sigma\vec{\nabla}_{\boldsymbol{\xi}} \qquad \partial_t \to \partial_t + \sigma\partial_\tau
\end{equation}
This should be understood as both the physical gradient $\vec{\nabla}$ and the time derivative $\partial_t$ being decomposed into fast and slow components \cite{hinch1991perturbation, nayfeh2008perturbation, bender1999advanced}. Accordingly, the first-order perturbations in the fluid amplitude $\delta$ are expanded as $n_1(\mathbf{x},t) = n_1^{(0)}(\mathbf{x},t; \boldsymbol{\xi}, \tau) + \sigma n_1^{h}(\mathbf{x},t; \boldsymbol{\xi}, \tau)$, etc. We substitute these fields into the system \eqref{eq continuidade ordem 1 n constante sistema 3} and \eqref{eq euler ordem 0 n constante sistema 3}, as well as the perturbed Gauss' law \eqref{lei de gauss plasmons} at first order, with the observation that $\mathbf{v}_0$ is a function of the slow variables only $(\boldsymbol{\xi}, \tau)$, while $n_0$ is taken to be a constant throughout, acquiring, at, zeroth order in $\sigma$,
\begin{numcases}
         {}\sigma\vec{\nabla}_{\boldsymbol{\xi}}\cdot\mathbf{v}_0 = 0 \label{velocidade fast}
         \\
         {}\left[\partial_t  + \left(\mathbf{v}_{0} \cdot \vec{\nabla}_\mathbf{x}\right)\right]n_1^{(0)} + n_0 \vec{\nabla}_\mathbf{x}\cdot \mathbf{v}_{1}^{(0)} = 0  , \label{eq continuidade fast}
        \\
		  \left[\partial_{t} +
       \left(\mathbf{v}_{0} \cdot \vec{\nabla}_\mathbf{x}\right) \right]\mathbf{v}_{1}^{(0)} 
       = - c_s^2 \vec{\nabla}_\mathbf{x}\left(\frac{n_1^{(0)}}{n_0}\right) - \frac{e}{m}\mathbf{E}_{1}^{(0)} , \label{eq euler fast}
       \\
       \vec{\nabla}_\mathbf{x} \cdot \mathbf{E}_1^{(0)} = -\frac{e}{\varepsilon}n_1^{(0)} . \label{eq campo eletrico fast}
\end{numcases}
Thus, at leading (fast) order, the material derivative does not contribute an advective term. At order $\sigma^1$ we obtain a system of slow secular equations, which we ignore, at present, to focus exclusively on the fast dynamics and their consequences.
Thus, taking the (fast) divergence of equation \eqref{eq euler fast}, we obtain
 \begin{equation}
 \label{equacao de euler div}
     n_0\left[\partial_{t} +
        \left(\mathbf{v}_{0} \cdot \vec{\nabla}_\mathbf{x}\right)\right] \nabla_\mathbf{x}^2\psi_{1}^{(0)} 
       = -c_s^2 \nabla_\mathbf{x}^2 n_1^{(0)} + \frac{q_e^2}{m_e}\frac{n_0 }{\varepsilon} n_1^{(0)} ,
\end{equation}
where we used the fact that, since $\mathbf{v}_0 = \mathbf{v}_0(\boldsymbol{\xi},t)$, the commutation relation $\left[\vec{\nabla}_\mathbf{x}D_t, D_t \vec{\nabla}_\mathbf{x}\right] = 0$ holds, where $D_t = \left(\partial_{t}+ \mathbf{v}_{0} \cdot \vec{\nabla}\right)$ is the material (or convective) derivative. 
Going further, we notice if we cast the continuity equation \eqref{eq continuidade fast} into the form
\begin{equation}
	 \nabla_\mathbf{x}^2 \psi_{1}^{(0)} = - \frac{1}{n_{0}}\left(\partial_{t}+ \mathbf{v}_{0} \cdot \vec{\nabla}_\mathbf{x}\right) n_{1}^{(0)} 
\end{equation}
and define the plasmon frequency
\begin{equation}\label{frequencia dos plasmons}
    \omega_p^2 = \frac{q_e^2}{m_e}\frac{n_0 }{\varepsilon} ,
\end{equation}
we can substitute these into equation \eqref{equacao de euler div} to obtain
\begin{equation}\label{equacao de klein-gordon plasmons}
    - \left[\partial_{t}+ \mathbf{v}_{0}\left(\tau,\boldsymbol{\xi}\right) \cdot \vec{\nabla}_\mathbf{x}\right]^2 n_{1}^{(0)}  + c_s^2 \nabla_\mathbf{x}^2 n_1^{(0)} - \omega_p^2 n_1^{(0)}  = 0.
\end{equation}
This is a real Klein--Gordon equation governing perturbations in an incompressible electron gas whose background velocity varies slowly over a wavelength; that is, it is the wave equation describing plasmons in an effective spacetime with metric signature $(-+++)$. A remarkable feature of this equation is the presence of a mass term, which distinguishes it from most wave equations encountered in most analogue gravity models. This mass term arises solely from the fact that the fluid is electrically charged. From this point onward, for notational simplicity, we shall omit the slow variables symbols $(\tau,\boldsymbol{\xi})$, the subscript denoting the fast spatial variable $\mathbf{x}$ and the superscript $(0)$ identifying the leading order perturbation, thereby facilitating the derivation of the effective metric tensor.

\subsection{The effective metric tensor}\label{subsec:tensor metrico}
In the subsequent analysis, we expand the material derivative in equation \eqref{equacao de klein-gordon plasmons}, following \cite{visser1998}. It yields
\begin{equation}
   -\Big[ \partial_t^2 n_1 + \partial_t(\mathbf{v}_0\cdot\vec{\nabla} n_1) + (\mathbf{v}_0\cdot\vec{\nabla})\partial_t n_1 + (\mathbf{v}_0\cdot\nabla)^2 n_1 \Big] + c_s^2 \nabla^2 n_1 - \omega_p^2 n_1 = 0 . \label{equacao de klein-gordon para os plasmons aberta}
\end{equation}
%
Since the Klein--Gordon equation for a scalar field $\varphi$ can be written as
\begin{equation} \label{equacao de klein-gordon generica}
\partial_{\mu}\left(f^{\mu\nu} \partial_{\nu}\varphi\right) - \omega^2 \varphi = 0 \quad \rightarrow f^{\mu\nu} \partial_{\mu}\partial_\nu\varphi +   \partial_{\mu} f^{\mu\nu} \partial_{\nu}\varphi - \omega^2\varphi = 0,
\end{equation}
we can compare equations \eqref{equacao de klein-gordon para os plasmons aberta} e \eqref{equacao de klein-gordon generica}, arriving at
\begin{align}
    f^{tt} = -1, \quad f^{ij} = c_s^2 - v_0^i v_0^j , \quad f^{it} = f^{ti} = -v_0^i .
\end{align}
Therefore,
\begin{equation}
    f^{\mu\nu} = 
    \begin{pmatrix}
        -1  & \vdots & -\mathbf{v}_0^j \\
         \hdots &  & \hdots\hdots\hdots \\
        -\mathbf{v}_0^i  & \vdots & c_s^2\delta^{ij} - v_0^i v_0^j 
    \end{pmatrix} .
\end{equation}
Identifying 
\begin{equation}
    G^{\mu\nu} \sqrt{-G} = f^{\mu\nu},
\end{equation}
it is simple to calculate
\begin{equation}
    \det(f^{\mu\nu}) = G^{-1} (\sqrt{-G})^4 = G = -c_s^6 ,
\end{equation}
and therefore $\sqrt{-G} = \sqrt{c_s^6} = \pm c_s^3$. Accordingly, the covariant metric tensor can be readily computed, resulting in
\begin{equation}
  G_{\mu\nu} = c_s
    \begin{pmatrix}
        -\left(c_s^2 - \mathbf{v}_0^2 \right)  & \vdots & -\mathbf{v}_0^j \\
         \hdots &  & \hdots\hdots\hdots \\
        - \mathbf{v}_0^i & \vdots & \delta_{ij}
    \end{pmatrix} . \label{tensor metrico acustico}
\end{equation}
This tensor represents an acoustic metric proportional to the one originally derived by Unruh and Visser. It therefore provides a natural extension of most analogue gravity models studied to date, which have primarily focused on massless scalar fields propagating on acoustic black hole backgrounds. In the present framework, one can instead investigate massive scalar fields in such analogue spacetimes. This effective metric governs the evolution of bulk plasmons in a jellium. In the following section, we will show how this evolution is translated for surface plasmons located at a conductor/dielectric interface.

\section{Surface scalar plasmons on curved interfaces}\label{sec:da costa}
\begin{figure}[h]
    \centering
    \includegraphics[width=1\linewidth]{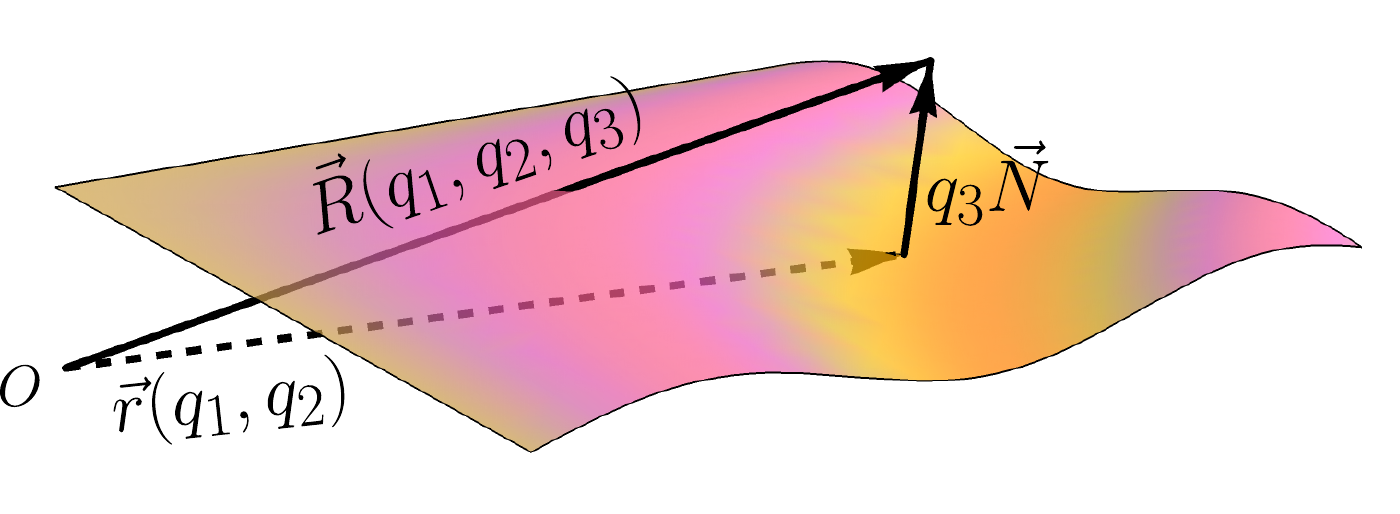}
    \caption{Parameterization of the Euclidean coordinate system.}
    \label{fig:placeholder}
\end{figure}

When investigating scalar fields, we come about a spatial Laplacian operator
	\begin{equation}
		\nabla^{2}=\frac{1}{\sqrt{G}} \partial_{i}\left(\sqrt{G} G^{i j} \partial_{j}\right).
	\end{equation}
Applying it to the scalar function $n(t,\mathbf{q})$ in three spatial dimensions, we get
\begin{equation}
\begin{aligned}
    \nabla^{2} n &= \frac{1}{\sqrt{G}} \partial_{i}\left(\sqrt{G} G^{i j} \partial_{j} n\right)+\frac{1}{\sqrt{G}} \partial_{3}\left(\sqrt{G} \partial_{3} n\right) \\
                &= \frac{1}{\sqrt{G}}\left[\partial_{i}\left(\sqrt{G} G^{i j} \partial_{j} n\right)\right]+\partial_{3} \ln \sqrt{G} \partial_{3} n+\partial_{3}^{2} n
\end{aligned}
\end{equation}
Consequently, the Klein--Gordon \eqref{equacao de klein-gordon plasmons} can be rewritten as
\begin{multline}
    -\left[\partial_{t} +
        \left(\mathbf{v}_{0} \cdot \vec{\nabla}\right)\right]^2 n_{1}
        + \frac{c_s^2}{\sqrt{G}}\left[\partial_{i}\left(\sqrt{G} G^{i j} \partial_{j} n_1 \right)\right] \\
        + c_s^2\left(\partial_{3} \ln \sqrt{G} \partial_{3} n_1 +\partial_{3}^{2} n_1\right)
        - \omega_p^2 n_1  = 0.
\end{multline}
Let us introduce a coordinate system to describe a point at a distance $|\mathbf{R}|$ from the origin and at a distance $q_3$ from a curvilinear object embedded in the Euclidean spacetime. We can parameterize this point as $\mathbf{R}(q_1,q_2,q_3) = \mathbf{r}(q_1,q_2) + q_3\mathbf{N}(q_1,q_2)$, as illustrated in Figure \ref{fig:placeholder}. 
The Euclidean spacetime can be described, in this parameterization, by the metric tensor $G_{i j} = \partial_i \mathbf{R} \cdot \partial_j \mathbf{R}$. Then, an infinitesimal volume $dV$ is 
\begin{equation}
    \begin{aligned}
     d V & =\left(\frac{\partial \vec{R}}{\partial q_{1}} \times \frac{\partial \vec{R}}{\partial q_{2}} \cdot \frac{\partial \vec{R}}{\partial q_{3}}\right) d q_{1} d q_{2} d q_{3} 
     \\
    & = \left(\sum_{j=1}^{2}\left(\delta_{1 j}+q_{3} a_{1 j}\right) \frac{\partial \vec{r}}{\partial q_{j}} \times \sum_{k=1}^{2}\left(\delta_{2 k}+q_{3} a_{2 k}\right) \frac{\partial \vec{r}}{\partial q_{k}}\right) \cdot \mathbf{N}(q_{1}, q_{2}) \ d q_{1} d q_{2} d q_{3} 
    \\
    & =\left[\left(1+q_{3} a_{11}\right)\left(1+q_{3} a_{22}\right)-\left(q_{3}^{2} a_{12} a_{21}\right)\right] \left(\frac{\partial \vec{r}}{\partial q_{1}} \times \frac{\partial \vec{r}}{\partial q_{2}} \right) \cdot \mathbf{N}(q_{1}, q_{2}) \ d q_{1} d q_{2} d q_{3} ,
    \end{aligned}
\end{equation}
where $a_{ij}$ is the shape operator \cite{tese-renato,tese-matheus}. Moreover, we identify the mean curvature $H$ and the gaussian curvature $K$ as
\begin{equation}
    H^2 = \left(
\frac{\text{Tr}\left(a_{ij}\right)}{2}\right)^2 \qquad \text{and} \qquad K = \det\left(a_{i j}\right). 
\end{equation}

It is well known that, under this description, one can define a surface metric tensor $g^{\mu\nu}$ together with the auxiliary function $f(q_1,q_2,q_3) = \left(1-2 H q_{3}+K q_{3}^{2}\right)$, written in terms of the mean curvature $H$ and Gaussian curvature $K$ of an arbitrary smooth surface, such that 
\begin{equation}
    \sqrt{G} = f(q_1,q_2,q_3) \sqrt{g} = \left(1-2 H q_{3}+K q_{3}^{2}\right)\sqrt{g},
\end{equation}
where $\sqrt{g}$ is independent of $q_3$. As a result, we get
\begin{equation}
    \partial_3 \ln \sqrt{G} = \partial_3 \ln f(q_1,q_2,q_3) + \partial_3 \ln \sqrt{g} = \frac{-2H + 2K q_3}{1-2 H q_{3}+K q_{3}^{2}}
\end{equation}
Therefore, substituting this into equation \eqref{equacao de klein-gordon plasmons}, we obtain
\begin{multline}\label{equacao de klein-gordon plasmons com as curvaturas}
    -\left[\partial_{t} +
        \left(\mathbf{v}_{0} \cdot \vec{\nabla}\right)\right]^2 n_{1}
        + \frac{c_s^2}{\sqrt{g}}\left[\partial_{i}\left(\sqrt{g} g^{ij} \partial_{j} n_1\right)\right] 
        \\
        + c_s^2\left[\left(\frac{-2H + 2K q_3}{1-2 H q_{3}+K q_{3}^{2}}\right) \partial_{3} n_1+\partial_{3}^{2} n_1\right] 
        - \omega_p^2 n_1  = 0.
\end{multline}
Now, in opposition to da Costa's approach, we do not impose an external confining potential followed by an artificial limiting procedure. 
Instead, we propose a physically motivated boundary condition: we assume that the surface plasmon is localized near the interface with an exponential decay on each side. This ansatz satisfies a Robin boundary condition at the interface,
\begin{equation}
    \partial_3 n_1 = -\alpha n_1 ,
\end{equation}
where $\alpha>0$ is the inverse penetration depth on the dielectric side ($q_3>0$). Using the convention that the conductor is medium $1$ and the dielectric is medium $2$, we require continuity of the density and of the normal component of the electric displacement at the interface:
\begin{equation}
    n_1^{(1)} = n_1^{(2)}, \qquad
    \epsilon_1 \partial_3 n_1^{(1)} = \epsilon_2 \partial_3 n_1^{(2)}
    \quad \text{at } q_3=0 .
\end{equation}
Thus we take
\begin{equation}
    n_1(t,q_1,q_2,q_3) =
    \begin{cases}
        A\eta(t,q_1,q_2) e^{-\alpha_2 q_3}, & q_3>0,\\[4pt]
        B\eta(t,q_1,q_2) e^{\alpha_1 q_3}, & q_3<0 .
    \end{cases}
\end{equation}
The continuity of the density gives $A=B$, while the continuity of the normal derivative gives $-\alpha_2 \epsilon_2 A = \alpha_1 \epsilon_1 B$, so that
\begin{equation}
    \alpha_1 \epsilon_1 + \alpha_2 \epsilon_2 = 0 .
\end{equation}
Imposing the Robin condition on the $q_3>0$ side yields $\alpha_2 = \alpha$, hence $\alpha_1 = -\alpha \epsilon_2/\epsilon_1$. Therefore
\begin{equation}
    n_1(t,q_1,q_2,q_3) =
    \begin{cases}
        A\eta(t,q_1,q_2) e^{-\alpha q_3}, & q_3>0,\\[4pt]
        A\eta(t,q_1,q_2) e^{-\alpha(\epsilon_2/\epsilon_1) q_3}, & q_3<0 .
    \end{cases}
\end{equation}
Substituting this result into the Klein--Gordon equation 
\eqref{equacao de klein-gordon plasmons com as curvaturas}, it follows that
\begin{multline}
   \Bigg\{ -\left[\partial_{t} +
        \left(\mathbf{v}_{0} \cdot \vec{\nabla}\right)\right]^2 
        + \frac{c_s^2}{\sqrt{g}}\left[\partial_{i}\left(\sqrt{g} g^{ij} \partial_{j} \right)\right] 
        \\
        + c_s^2\left[\left(\frac{-2H + 2K q_3}{1-2 H q_{3}+K q_{3}^{2}}\right) \partial_{3} +\partial_{3}^{2} \right] 
        - \omega_p^2 \Bigg\} n_1(t,q_1,q_2, q_3)  = 0.
\end{multline}
Calculating the derivatives and integrating across the interface (details are available in the Appendix \ref{apendice:integral}), we get
%
%
%
\begin{equation}\label{equacao de klein-gordon plasmons superficial} 
      -\left[\partial_{t} +
      \left(\mathbf{v}_{0} \cdot \vec{\nabla}\right)\right]^2 \eta
        + \frac{c_s^2 }{\sqrt{g}}\partial_{i}\left(\sqrt{g} g^{ij} \partial_{j} \eta \right)
        + \left[V_\text{geom}(q_1,q_2)   
        - \omega_\text{eff}^2 \right]\eta  = 0,
\end{equation}
where we identify
\begin{equation}\label{geometric potential}
   V_\text{geom}(q_1,q_2) = c_s^2\left(4H^2 - 2K\right)
\end{equation}
as a geometric potential and
\begin{equation}
    \label{massa efetiva}
    \omega_\text{eff}^2 = c_s^2\alpha_1^2 \frac{\epsilon_1}{\epsilon_2} +\omega_p^2 .
\end{equation}
as an effective frequency gap, which is proportional to an effective mass. The meaning of the parameters $\alpha_{1,2}$ is that of the attenuation coefficient, or equivalently, as the inverse of the penetration depth of the surface density
\begin{equation}
    d_{1,2} = \frac{1}{\alpha_{1,2}}.
\end{equation}
The larger the value of $\alpha_{1,2}$, the smaller the penetration depth of the wave. Equivalently, $\alpha_{1,2} = 0$ means that the wave is not attenuated, $\alpha_{1,2}>0$ corresponds to attenuation, and $\alpha_{1,2} < 0$ represents amplitude gain. 

We shall assume that the penetration depth is much smaller than the local radius of curvature of the surface. In terms of the mean curvature, this assumption can be written as
\begin{equation}
    d_{1,2} \ll H^{-1},
\end{equation}
Since the density is apreciable ony for $|q_3|\lesssim d_{1,2}$, we see that
\begin{equation}
    |q_3| \lesssim d_{1,2} \ll H^{-1}.
\end{equation}
Thus, if we consider a material with a large value of $\alpha_{1,2}$, the density $n_1$ will be appreciably different from zero only for very small values of $q_3$, that is, when $q_3 \ll H^{-1}$.
The exponential profile adopted here should be regarded as the simplest phenomenological description of a localized scalar surface plasmon. More sophisticated microscopic models may predict different decay laws, but the exponential constitutes the leading-order description of a mode with finite penetration depth.
%
\subsection{A new metric tensor}
Much like the case for finding the metric tensor in equation \eqref{equacao de klein-gordon plasmons}, we expand \eqref{equacao de klein-gordon plasmons superficial} as
\begin{multline}
   -\Big[ \partial_t^2 \eta + \partial_t(\mathbf{v}_0\cdot\vec{\nabla} \eta) + (\mathbf{v}_0\cdot\vec{\nabla})\partial_t \eta + (\mathbf{v}_0\cdot\nabla)^2 \eta \Big] + \frac{c_s^2 }{\sqrt{g}}\left[\partial_{i}\left(\sqrt{g} g^{ij} \partial_{j} \eta \right)\right] 
   \\
   + \left[ V_\text{geo}(q_1,q_2) - \omega_p^2 \right] \eta = 0 . \label{equacao de klein-gordon para os plasmons superficiais aberta}
\end{multline}
Since the Klein--Gordon equation for a scalar field $\varphi$ can be written as
\begin{equation} \label{equacao de klein-gordon generica 2}
\frac{1}{\sqrt{-\gamma}}\partial_{\mu}\left(\sqrt{-\gamma}\gamma^{\mu\nu} \partial_{\nu}\varphi\right) - \omega^2 \varphi = 0 \quad \rightarrow \quad \gamma^{\mu\nu} \partial_{\mu}\partial_\nu\varphi +   \partial_{\mu}\left(\sqrt{-\gamma} \gamma^{\mu\nu}\right) \partial_{\nu}\varphi - \omega^2\varphi = 0,
\end{equation}
we can compare equations \eqref{equacao de klein-gordon para os plasmons superficiais aberta} e \eqref{equacao de klein-gordon generica 2}, arriving at
\begin{align}
    \gamma^{tt} = -1, \quad \gamma^{ij} = c_s^2 g^{ij} - v_0^i v_0^j\mathbb{I} , \quad \gamma^{it} = \gamma^{ti} = -v_0^i .
\end{align}
Therefore,
\begin{equation}
    \gamma^{\mu\nu} = 
    \begin{pmatrix}
        -1  & \vdots & -\mathbf{v}_0^j \\
         \hdots &  & \hdots\hdots\hdots \\
        -\mathbf{v}_0^i  & \vdots & c_s^2 g^{ij} - v_0^i v_0^j
    \end{pmatrix} .
    \label{tensor metrico acustico superficial}
\end{equation}
and therefore $\sqrt{-\gamma} = \pm c_s^2/\sqrt{g}$, where $\gamma^{\mu\nu}$ is the new metric tensor in $2+1$ dimensions which allows us to cast our Klein--Gordon equation \eqref{equacao de klein-gordon plasmons superficial} in the standard form
\begin{equation}\label{equacao de klein-gordon plasmons superficial 2}
    \frac{1}{\sqrt{-\gamma}}\partial_{\mu}\left(\sqrt{-\gamma}\gamma^{\mu\nu} \partial_{\nu}\varphi\right) + \left[V(t,\mathbf{r}) - m^2 \right]\varphi = 0.
\end{equation}
This result echoes the well--known effective metric for gravity waves in water, proposed by Sch\"{u}tzhold and Unruh \cite{unruh-shallow}. Now, we can modify the effective metric not only by manipulating the background drift velocity field $\mathbf{v}_0$ using an external potential (remember equation \eqref{eq euler ordem 0 n constante sistema 3}) but also by engineering a curved interface described by the superficial metric tensor $g^{i j}$. Even in the absence of a drift velocity, one may still have a singular effective spacetime, provided that $g^{i j}$ represents a curved surface with a singularity.
\section{Physical Interpretation}\label{sec:interpretacao}
The superficial Klein--Gordon equation \eqref{equacao de klein-gordon plasmons superficial} possesses a striking feature: the perturbations in the electron fluid are subjected to a geometrically induced potential, represented by equation \eqref{geometric potential}. It follows that, depending on the macroscopic parameters $\alpha_{1,2},\epsilon_{1,2}$ and on the geometry of the interface, one may find a localized gapless (or massless) SSP if
\begin{equation}
      V_\text{geom}(q_1,q_2)  =  \omega_\text{eff}^2 .
\end{equation}
Thus, this system is adequate to investigate scenarios with position--dependent mass, dictated by geometry. Moreover, one finds that even in the most trivial of cases, the plane, where both mean and gaussian curvatures are zero, that the geometric potential does vanish, but there remains an effective mass correction, or a gap renormalization, yielding a new frequency $\omega_\text{eff}$ as defined by \eqref{massa efetiva}.
As an example, consider the interface of an ideal cylindrical conductor, described by $\mathbf{r} = (x,y,\sin kx)$, with $-L < x < L$. The gaussian curvature is $K = 0$ and the mean curvature can be calculated as
\begin{equation}
    H = -\frac{k^2 \sin (k x)}{2 \left[1 + k^2 \cos ^2(k x)\right]^{3/2}} ,
\end{equation}
where $k$ is chosen with the boundary condition $\sin kL = 0$, meaning $k = n\pi/L$, for $n$ integer. 
\begin{figure}[h]
    \centering
    \includegraphics[width=0.75\linewidth]{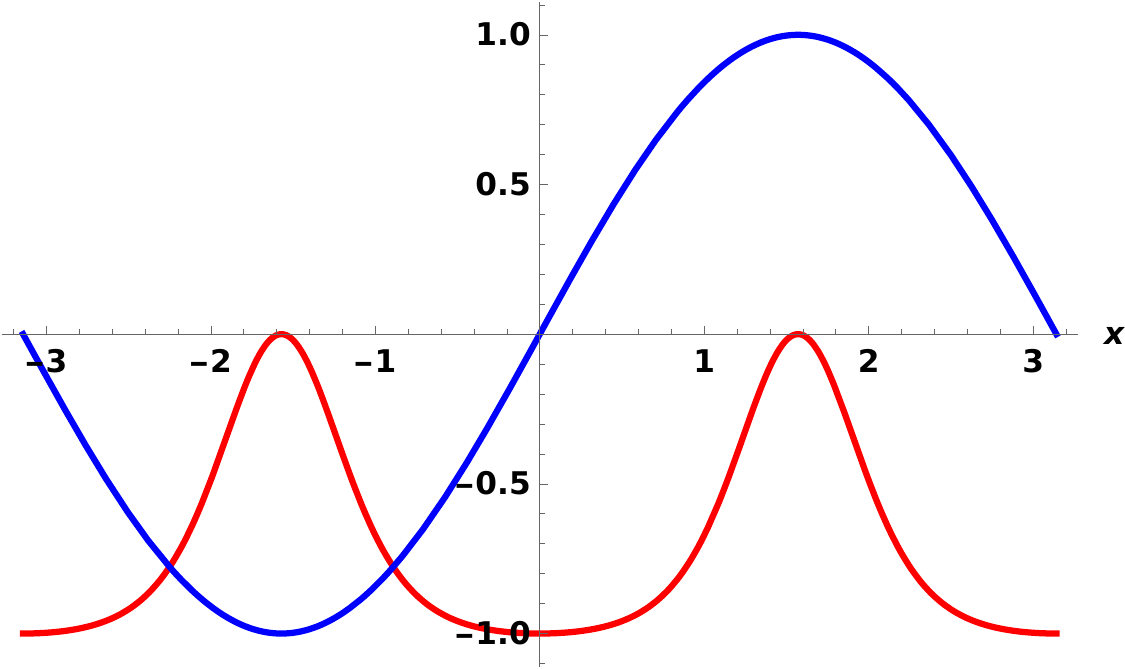}
    \caption{Sinusoidal cylindrical surface and potential.}
    \label{fig:seno}
\end{figure}
The condition for massless propagation, for background flow $\mathbf{v}_0 = 0$, is
\begin{equation}\label{potencial geometrico seno}
-1  +  \frac{k^4 \sin^2 (k x)}{ \left[1 + k^2 \cos ^2(k x)\right]^{3}}  = 0,
\end{equation}
where we consider $\epsilon_1/\epsilon_2$ real and use units such that
\begin{equation*}
    \alpha_1^2 \frac{\epsilon_1}{\epsilon_2} + \frac{\omega_p^2}{c_s^2} = 1 .
\end{equation*}
Choosing $L = \pi$ and $n = 1$, we find that the only two possible locations where massless perturbations occur are at $x = \pm \pi/2$, coinciding with the maxima of our effective potential $V_\text{eff}(x) =  V_\text{geom}(x)  - \omega_\text{eff}^2$. The minimum is located at $x = 0$ and is exactly where the curvature of our idealized cylinder vanishes and where the perturbations are more likely to be propagating. As illustrated in Figure \ref{fig:seno}, we see the location of our maxima and minima relative to the profile of our interface. Nonetheless, one can conceive other examples of carefully constructed mediums, such as a tube with circular cross--section, whereby the choice of it radius dictates the energy gap (or the effective mass) of the plasmons.  
Therefore, given a geometry, one may readly analyse other important regimes, such as 
\begin{equation*}
         V_\text{geom}(q_1,q_2)  < \omega_\text{eff}^2 ,
\end{equation*}
which may be related to a tachyonic--like \cite{tachyon} instability, and
\begin{equation*}
        c_s^2  V_\text{geom}(q_1,q_2)  > \omega_\text{eff}^2  ,
\end{equation*}
manipulating an energy gap for the plasmons. 
The geometry of the interface, therefore, plays a crucial role in determining the collective dynamics of the conduction electrons. 
We may we determine the plasmonic modes directly from the wave equation \eqref{equacao de klein-gordon plasmons superficial} governing the SSP evolution, leading to an eigenvalue problem for the allowed frequencies. 

While these preliminary results may have implications in plasmon transport theory, which lay beyond the scope of this study, we would like to present a different kind of application, one which bridges two seemingly unrelated areas: condensed matter physics and general relativity.
\section{An application in analogue gravity}\label{sec:analogue}
It is clear from equation \eqref{equacao de klein-gordon plasmons superficial} that we can use it to understand the dynamics of a massive scalar field in a curved spacetime, with a position--dependent mass. As a motivation, we recall that the evolution of a massive field at a spacetime under a nonminimal coupling with the metric can described by a wave equation, established in the literature  \cite{faraoni-nonminimal,nonminimal2,nonminimal3,nonminimal4} as
\begin{equation}
    \Box\phi - \xi R \phi - \frac{dV}{d\phi} = 0,
\end{equation}
where $\Box$ is the d'Alambertian operator, $R$ is the Ricci curvature of spacetime, $V(\phi)$ is a scalar potential, and $\xi$ is the coupling strength (commonly found as $\xi = 1/6$ for a conformal coupling). If we assume the units $c = \hbar = 1$ and $V(\phi) = m^2\phi^2/2$, we find that
\begin{equation}
    \Box\phi - \left(\xi R + m^2\right)\phi = 0.
\end{equation}
It is a simple matter to compare the last expression with equation \eqref{tensor metrico acustico superficial} to find that one is very similar, and in fact equal to the other, provided that we consider this in a fixed spatial slice, reducing it to $2+1$ dimensions, and that we make the identifications
\begin{equation}\label{mapping}
    -\left(\xi R + m^2\right) \Leftrightarrow  V_\text{geom} - \omega_\text{eff}^2
\end{equation}
%
The mapping above is established with conjunction to the effective metric \eqref{tensor metrico acustico superficial}, which can be modeled at will, making equation \eqref{equacao de klein-gordon plasmons superficial} an addition to the analogue gravity programme with possible applications to the simulation of the $f(R)$ extensions of General Relativity. As an illustrative example, we consider a plane conductor--dielectric interface, where our conductor has permittivity $\epsilon_1$ and the dielectric has permittivity $\epsilon_2$.

\begin{figure}[h]
    \centering
    \includegraphics[width=1\linewidth]{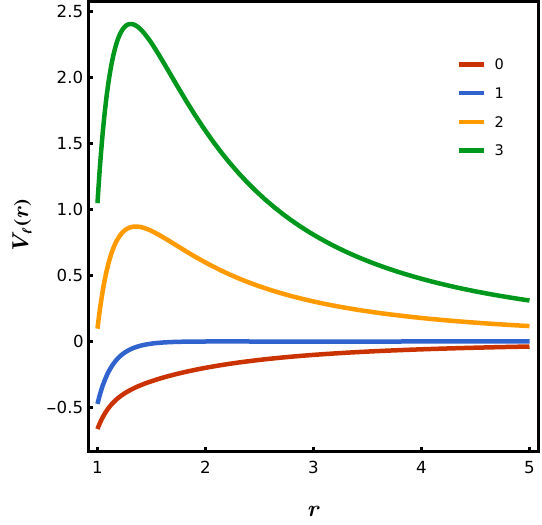}
    \caption{Effective potential for $\ell = 0,1,2,3$. We use $c_s = 1, \omega_p = 1$ and $A = 0.9$.}
    \label{fig:potenciais}
\end{figure}

The plane is both the most trivial and most illustrative example of a geometry that can exhibit a sonic horizon. Because the planar geometry has no curvatures, the geometric potential will be zero and we will still have an effective mass due to the properties of the interface. In the cylindrical representation for a plane at height $z_0$, a point on this plane have the position vector $\mathbf{x} = (r\cos\phi,r\sin\phi,z_0)$ , which defines the surface metric as
\begin{equation}
    g_{ij} = 
    \begin{pmatrix}
        1 & 0 \\
        0 & r^2
    \end{pmatrix} .
\end{equation}
The incompressibility condition in cylindrical coordinates reveals that one possibility for the background velocity field is
\begin{equation}\label{background}
    \frac{1}{r}\partial_i\left(r v_0^i\right) = 0 \quad \rightarrow \quad \mathbf{v}_0 = -\frac{A}{r}\hat{\mathbf{r}},
\end{equation}
which has a corresponding external electric field given by equation \eqref{eq euler ordem 0 n constante sistema 3}. This represents the simplest possible velocity field with $\mathbf{v}_0^\phi = 0$, where we choose $A$ as a positive integration constant, directly related to the flux through a circle of radius $r$. With the help of the function
\begin{equation}
    f(r) = c_s^2 - \frac{A^2}{r^2},
\end{equation}
we locate the sonic horizon at $f(r_h) = 0$, that is, at $r_h = |A|/c_s$, and because the curvatures are zero, we use the \textit{ansatz} $\eta_\ell(t,r,\phi) = e^{-i\omega t} e^{i \ell \phi} S(r)$ to find
\begin{equation}\label{equacao radial} 
 \frac{d^2 S}{dr^2} +  P(r) \frac{d S}{dr}  + Q(r)S(r)  = 0,
\end{equation}
where we use
\begin{align}
        P(r) &= \frac{A^2 - 2 i A r^2 \omega +c_s^2 r^2}{c_s^2 r^3-A^2 r},
        \\
        Q(r) &= \frac{c_s^2\ell^2 - \Omega^2 r^2}{A^2-c_s^2 r^2},
        \\
        \Omega^2 &= \omega^2 - \omega_\text{eff}^2  . \label{def Omega maiusculo}
\end{align}
Before working out the exact solution for this equation, it is helpful to analyse its behavior as it approaches the singularities located at $r = r_h$ and at $r \to \infty$. To do so, we introduce the change of variables 
\begin{equation}
    S(r) = \sqrt{f(r)}\exp{\left[-\frac{1}{2}\int P(r) dr\right]}U(r).
\end{equation}
Now, we define a tortoise--like coordinate $r_*(r)$ via
\begin{equation}
    \frac{d r_*}{dr} = \frac{1}{f(r)} \quad \rightarrow \quad r_*(r) = \frac{1}{c_s^2} \left(r-\frac{r_h}{2} \ln\left|\frac{1 + r/r_h}{1 - r/r_h}\right|\right)  ,
\end{equation}
With this coordinate, we map the interval $(r_h,\infty)$ into $(-\infty,\infty)$. More specifically, we see that, asymptotically,
\begin{equation}\label{tartaruga assintotica}
    r_* \to \frac{1}{c_s^2}
    \begin{cases}
          r, \quad &r \to \infty 
        \\
        \frac{r_h}{2 } \ln\left(r/r_h - 1\right), \quad &r \to r_h 
    \end{cases} .
\end{equation}
This variable transmutes equation \eqref{equacao radial} into
\begin{equation}\label{eq de onda tartaruga}
   \frac{d^2 U}{d r^{*2}} + \left[  c_s^2 \Omega^2 - V_\text{eff}(r)\right] U(r_*) = 0.
\end{equation}
We call $V_\text{eff}(r)$ an effective potential, to which our massive plasmon is subject to, and it has the form
\begin{equation}\label{potencial efetivo}
   V_\text{eff}(r) = f(r)\left[\frac{A^2 \left(\Omega ^2 - \omega ^2\right)}{r^2 f(r)}+\frac{5 A^2}{4 r^4}-\frac{c_s^2(1 - 4\ell^2)}{4 r^2}\right].
\end{equation}
We plot the potential for the first few values of $\ell$ in Figure \ref{fig:potenciais}, where we see that for $\ell > 1$ not only the potential is now positive but also there appears an angular momentum barrier, which shifts the locations of the eigenvalues in the complex plane. 
The asymptotics of \eqref{potencial efetivo} are easy to check, being
\begin{equation}
    V_\text{eff}(r) \to 
    \begin{cases}
    0 , &\quad r \to \infty 
    \\
            c_s^2 (\Omega^2 - \omega^2), &\quad r \to r_h
    \end{cases} .
\end{equation}
allowing us to write the solution of the wave equation \eqref{eq de onda tartaruga} as
\begin{equation}\label{solucao tartaruga}
    U(r_*) = 
    \begin{cases}
        a_1 e^{-i c_s\Omega r_*} + a_2 e^{i c_s\Omega r_*}, \quad &r \to \infty \quad (r_* \to \infty)
        \\
         b_1 e^{-i c_s\omega r_*} + b_2 e^{i c_s\omega  r_*}, \quad &r \to r_h \quad (r_* \to -\infty) 
    \end{cases} .
\end{equation}
We take the positive branch for the eigenvalues, such that $\Omega \in \mathbb{R}_+$, $a_1$ represents the amplitude of an incoming wave from infinity
and $a_2$ represents the amplitude of an outgoing wave from infinity. Accordingly, $b_1$ is the amplitude of an incoming wave at the horizon and, in a similar manner, $b_2$ is the amplitude of an outgoing wave from the horizon. Additionally, imaginary $\Omega$ transforms $a_1$ into the amplitude of an exponential growing mode at infinity and
$a_2$ into the amplitude of an exponential decaying mode at infinity, if we take the positive imaginary branch. Because $\omega$ can also be complex, we find that there are three types of physical solutions \cite{alves} to our problem. First, quasibound modes, where $\eta \to 0$ at spatial infinity and $b_1 \neq 0$ at $r = r_h$. Second, quasinormal modes, where $\eta \propto e^{i c_s\Omega r_*}$ at infinity and again $b_1 \neq 0$ at $r = r_h$. In both cases, we take $a_1 \to 0$ in order to match the physical requirements that the wave propagates to infinity (QNM) or decay at infinity (QBS). In general, we will write $\omega = \omega_R - i\omega_I$ and $\Omega = \Omega_R - i\Omega_I$, where $\Omega_I > 0$ identify QNMs and $\Omega_I < 0$ identify QBSs. The third case is of scattering modes, where the frequencies $\Omega$ are all real, meaning $\omega^2 > \omega_\text{eff}^2$, and we can choose the amplitude $a_1$ of the incoming wave from infinity.

To make the asymptotics of our wave transparent, we return to the radial equation \eqref{equacao radial} with the original coordinate $r$ using equation \eqref{tartaruga assintotica}, writing the asymptotic solutions as
\begin{equation}\label{solucao assintotica geral}
    S(r) \sim  
    \begin{cases}
        r^{ -1/2 + i r_h
   \omega /c_s} \left(a_1 e^{-i \Omega r/c_s} + a_2 e^{i \Omega r/c_s}\right) ,\quad &r \to \infty 
        \\
        b_1 + b_2 \left(\frac{r}{r_h}\right)^{i r_h\omega/ c_s}, \quad &r \to r_h  
    \end{cases} 
\end{equation}
where we have absorbed all constant factors into $a_{1,2}$ and $b_{1,2}$. Since we are using the convention $e^{-i\omega t}$, we know that $b_2$ must be zero for a physically meaningful solution, because at the horizon of the (analogue) black hole, there must be only incoming waves.
Therefore, we need to solve equation \eqref{equacao radial} and use equations \eqref{solucao assintotica geral} as the asymptotic behavior of our solutions in order to extract the eigenfrequencies $\omega$.
\subsection{The confluent Heun equation}

To solve our radial--like problem in equation \eqref{equacao radial} locally around the horizon $(r = r_h)$, we rescale our variable by defining $u = r^2/r_h^2$ and $S(r) = U(u)$. We use the \textit{ansatz}
\begin{equation}
    U(u) = u^B (u-1)^C e^{Du} H(u),
\end{equation}
shaping equation \eqref{equacao radial} into 
\begin{equation}\label{eq de Heun}
    \frac{d^2 H}{d u^2} + \left(\frac{\gamma'}{u} + \frac{\delta'}{u - 1} + \epsilon' \right)\frac{d H}{d u} + \left[\frac{\alpha' u - q}{u(u-1)}\right]H(z) = 0,
\end{equation}
This is a confluent Heun equation where the parameters are
%
\begin{equation}\label{parametros heun}
    \begin{aligned}
        \gamma ' &=  2B ,
        \\
        \delta' &= 2C + 1 - i r_h \omega/c_s,
        \\
        \epsilon' &= 0,
        \\
        \alpha' &= \frac{r_h^2 \Omega^2}{4 c_s^2},
        \\
        q &=   \frac{i r_h \omega}{c_s}B - (2
   C + 1)B + \frac{\ell^2}{4},
    \end{aligned}
\end{equation}
and we can choose any combination of the exponents  and $B = \{0, 1\}$ and $C = \{0, i r_h \omega / c_s\}$, while $D = 0$. Because we have the special case where $\epsilon' = 0$, we refer to this equation as a reduced confluent Heun equation (RCHE) \cite{Bartolomeu_2002,Bartolomeu_2005,Bartolomeu_2008,Bartolomeu_2015,Ishkhanyan_2005_Beta}. Consequently, based on the boundary conditions \eqref{solucao assintotica geral}, the correct exponents are $B = C = 0$ for the first solution and $C = ir_h \omega/c_s$ with $B = 0$ for the second solution. To solve equation \eqref{eq de Heun} locally, we use $z = 1-u$ and employ a Frobenius series centered at $z = 0$, that is, at the horizon, with the form
\begin{equation} 
    \label{frobenius}
    H(z) = \sum_{n=0}^\infty c_n z^{n+s} .
\end{equation}
We find an indicial equation $s(s-1+\gamma')=0$, whose solutions reveal that the characteristic exponent $s$ can take two values, namely, $s_1=0$ or $s_2=1-\gamma' $. For $s = s_1$, we obtain the three--term recurrence relation between the $c_n$ coefficients
\begin{equation}
    \label{3-term-recurrence-CHE}
c_{n+1} = A_n c_n + B_n c_{n-1},
\end{equation}
where
\begin{equation}
    \label{An-Bn}
A_{n} =  \frac{(n+s)(n+s-1+\gamma') - q}{(n+s+1)(n+s+\gamma')} , \ \quad B_n =  \frac{\alpha}{(n+s+1)(n+s+\gamma')}.
\end{equation}
The $c_n$ coefficients can be determined transforming the three-term recurrence relation \eqref{3-term-recurrence-CHE} into a tridiagonal $(n+1)\times (n+1)$ matrix using \eqref{An-Bn}. Now, we impose that the determinant
\begin{equation}
D_{n+1} = \left|
\begin{array}{cccccccc}
	A_0 & -1 & 0 & 0 & 0 & \cdots & \cdots & \cdots \\
	B_1 & A_1 & -1 & 0 & 0 & \cdots & \cdots & \cdots \\
	0 & B_2 & A_2 & -1 & 0 & \cdots & \cdots & \cdots \\
	\cdots & \cdots & \cdots & \cdots & \cdots & \cdots & \cdots & \cdots \\
	\cdots & \cdots & \cdots & \cdots & \cdots & \cdots & A_{n-1} & -1 \\
	0 & 0 & 0 & 0 & 0 & \cdots & B_n & A_n
\end{array} \right| 
\end{equation}
must vanish, i.e., $ \lim_{n\rightarrow \infty} D_n = 0$, to guarantee the existence of the solution of the homogeneous system. Thus, solving this equation gives us a list of eigenvalues $\omega$ that make the series \eqref{frobenius} convergent in the unit disk $|1 - r^2/r_h^2| < 1$. A similar situation happens when we use $s = s_2$. 
Then, it is simple to generate all coefficients using
\begin{equation}
    \label{cn} 
    c_n = D_n c_0,
\end{equation}
where $c_0=1$ initializes the sequence \cite{kristensson}. Finally, build the solution around the horizon as
\begin{equation}\label{solucao geral radial}
    H^{0}(u) = b_1 H^{0+}(u) + b_2 H^{0-}(u) .
\end{equation}
where $b_{1,2}$ are constants and
\begin{align}
   H^{0+}(u) &= \text{HeunC}\left(\frac{\ell^2}{4} -\frac{r_h^2 \Omega^2}{4 c_s^2}, -\frac{r_h^2 \Omega^2}{4 c_s^2}, 1 - i \frac{r_h \omega}{c_s}, 0, 0; 1 - u\right) ,
    \\
    H^{0-}(u) &= \left(1 - u\right)^{i r_h \omega/c_s} \text{HeunC}\left(\frac{\ell^2}{4} -\frac{r_h^2 \Omega^2}{4 c_s^2}, -\frac{r_h^2 \Omega^2}{4 c_s^2}, 1 + i \frac{r_h \omega}{c_s} , 0, 0; 1 - u\right) .
\end{align}
In this notation, $H^{Z_0\pm}$ means the solution is around a singularity located at $z = Z_0$ (either $0, 1, \infty$) and the $+$ sign means 'ingoing \textit{to}' the singularity, while the negative sign $-$ means 'outgoing \textit{from}' the singularity.
Now, we would like to connect our solution \eqref{frobenius}, which is valid in $0 < u < 2$, with some asymptotic solution around $u\to\infty$. To do so, we need to solve the confluent Heun equation \eqref{eq de Heun} asymptotically and approach this two--point connection problem. Here, we employ the method introduced by Fiziev \cite{Fiziev_2006}, which was originally done with the complete CHE ($\epsilon' \neq 0$) in the solution to the Regge--Wheeler equation. Because we have $\epsilon' = 0$, our solution will be of the subnormal type \cite{ince, olver-asymptotics,tese_trieste,Bonelli_2022}, written as
\begin{equation}
    \label{subnormal}
    H^{\infty\pm}(u) \rightsquigarrow  u^{\lambda} \exp\left(\pm 2\sqrt{-\alpha'}u^{1/2}\right)\sum_{n = 0}^\infty \frac{a_n^{\pm}}{u^{n/2}} ,
\end{equation}
where the positive and negative branches represent the two different solutions and
\begin{equation}
    \lambda = \frac{1}{4} - \frac{\gamma' + \delta'}{2}.
\end{equation}
This \textit{ansatz} generates a recurrence relation for the coefficients $a_k^{\infty\pm}$ in the form
\begin{equation}\label{relacao de recorrencia infinito}
    a_{n-1}^{\pm} = A_n^{\pm} a_{n}^{\pm} + B_n ^{\pm} a_{n + 1}^{\pm} + C_n^{\pm}a_{n + 2}^{\infty\pm},
\end{equation}
where $n \geq 1$ and 
\begin{equation}\label{coeficientes-infinito}
    \begin{aligned}
        A_n^{(\pm)} &=   \frac{\mp 4\sqrt{-\alpha'}( n + \delta')}{\left[2 n + 1 - 2(\gamma - \delta)\right]\left[2 n - 3 + 2(\gamma + \delta)\right]},
        \\
        \\
        B_n^{(\pm)} &=   -\frac{4(n + 2)(n+1) + 16(\alpha' - q) - 4\left[-3 + 2(\gamma+\delta)\right]\left[-1 + 2(\gamma+\delta)\right]}{\left[2 n + 1 - 2(\gamma - \delta)\right]\left[2 n - 3 + 2(\gamma + \delta)\right]},
        \\
        \\
        C_n^{(\pm)} &=   \frac{\pm 16 \sqrt{-\alpha'}(n + 2) }{\left[2 n + 1 - 2(\gamma - \delta)\right]\left[2 n - 3 + 2(\gamma + \delta)\right]}.
\end{aligned}
\end{equation}
The solutions are
\begin{align}
    \label{solucoes subnormais}
    H^{\infty +}(u) &\rightsquigarrow  u^{1/4 - (1 - i r_h \omega/c_s)/2} \exp\left(\frac{r_h}{c_s}i\sqrt{\omega^2 - \omega_\text{eff}^2} \ u^{1/2}\right)\sum_{n = 0}^\infty \frac{a_n^{(+)}}{u^{n/2}},
    \\
    H^{\infty -}(u) &\rightsquigarrow  u^{1/4 - (1 - i r_h \omega/c_s)/2 } \exp\left(- \frac{r_h}{c_s}i\sqrt{\omega^2 - \omega_\text{eff}^2} \ u^{1/2}\right)\sum_{n = 0}^\infty \frac{a_n^{(-)}}{u^{n/2}}. \label{solucao assintotica boa}
\end{align}
Following Fiziev, we can now write 
\begin{equation}
    H(u) = 
    \begin{cases}
        k_{1+} H^{0+}(u) + k_{1-} H^{0-}(u), \quad &0 < u < 2 
        \\
       k_{\infty+} H^{\infty+}(u) +  k_{\infty-} H^{\infty-}(u) \quad &1 < u < \infty
    \end{cases} .
\end{equation}
To analyse this problem and present numerical results, we employ a modified version of the method developed by Fiziev \cite{fiziev2006}. In his work, Fiziev explicitly solved the complete CHE, not the reduced case. In the next subsection, we present an adaptation of his method in order to adequately calculate the frequency eigenvalues that characterize our perturbations.
\subsection{The Reduced Fiziev Method}
The most correct way of calculating the eigenvalues of our physical system is to expand our solutions around a singular point as linear combinations of solutions around another singular point. This is a two--point boundary problem \cite{Fiziev_2006,motohashi}, and it can be solved using Wronskians to analytic continue our solutions, such that they can be evaluated in the domain of interest, which in our case is $(1,\infty)$. The coefficients of these expansions are directly related to wronskians and connect two different domains. This connection is made possible by boundary conditions at $u = 1$ and $u = \infty$, which impose linear dependence on two our solutions, thus forcing one wronskian determinant to be zero, from which we find the desired eigenvalues.

However, as this may be somewhat computationally demanding, Fiziev introduced a method for finding eigenvalues more easily, not by solving an equation involving a wronskian of two very complicated functions, but by exploring a workaround using the linear dependence between solutions. Because we need to apply the Fiziev method to the RCHE \eqref{eq de Heun}, we refer to this as the reduced Fiziev method (RFM). Here, one must write its solutions around the horizon as linear combinations of the solutions around infinity, such that
\begin{align}
        H^{0+}(u) &= \Gamma^{0+}_{\infty+} H^{\infty+}(u) +  \Gamma^{0+}_{\infty-} H^{\infty-}(u) \label{solucao in horizonte} 
        \\
        H^{0-}(u) &= \Gamma^{0-}_{\infty+} H^{\infty+}(u) +  \Gamma^{0-}_{\infty-} H^{\infty-}(u) \label{solucao out horizonte} 
\end{align} 
for transition coefficients $\Gamma^{X\pm}_{Y\pm}$, where $X$ represents a singular point and $Y \neq X$ represents the other singular point. The general solution can be expanded locally as
\begin{equation}\label{solucao geral local}
    H(u) = C_{0+}H^{0+}(u) + C_{0-}H^{0-}(u) = C_{\infty+}H^{\infty+}(u) + C_{\infty-}H^{\infty-}(u),
\end{equation}
where $C_{X\pm}$ are constants. From this last equation, one realizes, after substitution, that
\begin{equation}\label{coeficientes infinito e transicao}
    C_{\infty\pm} = C_{0+} \Gamma^{0+}_{\infty\pm} + C_{0-}\Gamma^{0-}_{\infty\pm}.
\end{equation}
At this point, we remember that, for quasinormal and quasibound states, the boundary conditions demand only incoming waves at the horizon and outgoing waves at infinity. Mathematically, this means imposing $C_{0-} = C_{\infty-} = 0$ in equation \eqref{solucao geral local}, which is the same as
\begin{equation}
    H^{0+}(u) \propto H^{\infty+}(u).
\end{equation}
The consequence of the last equation is clear: the Wronskian between the physically acceptable solutions will be 
\begin{equation}
    \mathcal{W}\left[H^{0+}(u_0), H^{\infty+}(u_0)\right] = H^{0+}(u_0) \frac{d}{du}H^{\infty+}(u_0) - H^{\infty+}(u_0) \frac{d}{du}H^{0+}(u_0)  \equiv 0
\end{equation}
which determines our eigenvalues for any point $u_0 \gg 1$. Instead of using this equation, however, we can using Fiziev's workaround, because there are not known closed form representations for derivatives of confluent Heun functions. What we do is calculate the absolute value of the constants $C_{\infty\pm}$ using the asymptotic behavior of the stationary general solution \eqref{solucao geral local}.
In general, given that $n \in \mathbb{Z}_+$ and $\lambda,\alpha', u \in \mathbb{C}$, we calculate
\begin{equation}
   \lim_{u \to \infty} \left|H^{\infty\pm}(u)\right| \sim \lim_{u \to \infty}\left|u^\lambda\exp\left(\pm 2\sqrt{-\alpha'}u^{1/2}\right)\sum_{n = 0}^\infty \frac{a_n^{\infty\pm}}{u^{n/2}}\right| \sim \left|u^\lambda\exp\left(\pm 2\sqrt{-\alpha'}u^{1/2}\right)\right|.
\end{equation}
Now we use the polar form for the complex variables $u$ and $\alpha'$, which we define as $u = u_r e^{i\theta}$, $\alpha' = \alpha'_r e^{i\phi}$, resulting in
\begin{multline}
    \lim_{u \to \infty} \left|H^{\infty\pm}(u)\right| \sim \left|u^\lambda\exp\left[\pm 2i\sqrt{\alpha'_r u_r}e^{i\left(\phi + \theta\right)/2}\right]\right| 
    \\
    \sim \left|u^\lambda\exp\left[\mp 2\sqrt{\alpha'_r u_r}\sin \left(\frac{\phi+ \theta}{2}\right)\right]\exp\left[\pm 2i\sqrt{\alpha'_r u_r}\cos\left(\frac{\phi + \theta}{2}\right) \right]\right|.
\end{multline}
The asymptotic behavior of our solution now amounts on the real exponential. Indeed, there are three cases to consider. First,
\begin{equation}
    \sin \left(\frac{\phi + \theta}{2}\right) < 0 \quad \rightarrow \quad (2m-1)\pi < \frac{\phi + \theta}{2} < 2m\pi,
\end{equation}
for integer $m$. From this, it is obvious that the argument of the exponential switches sign, making the previously $(-)$ decaying solution to grow up to infinity and the previously $(+)$ growing solution to decay to zero. This region of the complex plane is called a Stokes sector \cite{slavyanov, fiziev2006} for a given $m$. Second,
\begin{equation}
    \sin \left(\frac{\phi + \theta}{2}\right) > 0 \quad \rightarrow \quad 2m\pi < \frac{\phi + \theta}{2} < (2m+1)\pi,
\end{equation}
for integer $m$. In this case, we see that the argument of the exponential remains the same, which means that the solution either decays or blows up as $u_r$ grows to infinity. This region of the complex plane defines another Stokes sector \cite{slavyanov, fiziev2006} for a given $m$. The third and last case is
\begin{equation}\label{raio de stokes}
    \sin \left(\frac{\phi + \theta}{2}\right) = 0 \quad \rightarrow \quad  \frac{\phi + \theta}{2} = \pm m\pi,
\end{equation}
where an $m$ defines two lines (positive and negative branches). Each line is called a Stokes ray (or Stokes line), which separates two Stokes sectors. In this last case, we see that the solutions $\infty\pm$ evolve as
\begin{equation}
    \lim_{u \to \infty} \left|H^{\infty\pm}(u)\right| \sim \left|u^\lambda \exp\left[\pm 2i\sqrt{\alpha'_r u_r}(-1)^m \right]\right|.
\end{equation}
meaning that depending on the ray, defined by even or odd $m$, the solutions $\infty\pm$ will switch signs and limits. We can go one step further and define \cite{slavyanov} an anti--Stokes ray, which divides a Stokes sector in half, as
\begin{equation}\label{anti-raio de stokes}
    \frac{\phi + \theta}{2} = \left(\frac{1}{2} \pm m \right)\pi.
\end{equation}
On an anti--ray, we get
\begin{equation}
     \lim_{u \to \infty} \left|H^{\infty\pm}(u)\right| \sim  \left|u^\lambda\exp\left[\mp 2\sqrt{\alpha'_r u_r}(-1)^m \right]\right|,
\end{equation}
because $\cos\left[(m + 1/2)\pi\right] = 0$. On an anti--Stokes ray the solutions no longer oscillate, in contrast with the normal (or Thom\'{e}) solutions, either decaying or blowing up exponentially. The direction in which our solution decay exponentially fast corresponds to the steepest descent direction. Thus, following Fiziev \cite{fiziev2006}, the anti--Stokes ray is the optimal direction to compute the eigenvalues. All these conditions depend on the integer $m$, which, in a real calculation, is chosen \textit{a priori} such that we define a Stokes sector, two Stokes rays using \eqref{raio de stokes} and an anti--Stokes ray via \eqref{anti-raio de stokes}. In this work, we are interested in analysing the QNM boundary conditions. Thus, we use an anti--Stokes ray as in equation \eqref{anti-raio de stokes},choosing $m = 0$ to compute our radial--like eigenvalues in equation \eqref{equacao radial}, meaning
\begin{equation}
  \theta = \pi - \phi .
\end{equation}
Now that we know the optimal direction in the complex plane in which to calculate the eigenvalues, from equation \eqref{solucao geral local} we obtain the limits
\begin{align}
    |C_{\infty+}| &= \lim_{u\to \infty} \left|u^{-\lambda}\exp\left(+2i\sqrt{\alpha'}u^{1/2}\right)H(u)\right|,
    \\
    |C_{\infty-}| &= \lim_{u\to \infty} \left|u^{-\lambda}\exp\left(-2i\sqrt{\alpha'}u^{1/2}\right)H(u)\right|, \label{coeficiente infinito menos}
\end{align}
evaluated on a Stokes anti--ray. Because we selected $m = 0$, this enforces $|C_{\infty-}| \to 0$ and the consequence is that we must have the limit
\begin{equation}\label{limite heunc}
    \lim_{u\to\infty} \left|\text{HeunC}\left(q,\alpha', \gamma', \delta',\epsilon'; 1 - u_r\exp\left[i\left(\pi -\phi\right)\right]\right)\right| = 0,
\end{equation}
where $u = u_r e^{i\theta}$ and $\alpha' = \alpha'_r e^{i\phi} = X_{\alpha'}+ i Y_{\alpha'}$. Numerically, instead of calculating an actual limit, it is enough to evaluate this expression at $|u| = u_0 \gg 1$. This evaluation is what enables us to calculate what eigenvalues allow the local confluent Heun function $\text{HeunC}(\hdots ; u)$ to be evaluated outside of its usual domain, or in other words, we are constructing an analytic continuation of this function.
Finally, after solving this problem, we are able to write down the complete solution to the $\ell-$mode of our plasmonic dynamics as
\begin{equation}
    \eta_\ell(t,r,\phi) = e^{-i\omega t} e^{i \ell \phi}\text{HeunC}\left(\frac{\ell^2}{4} -\frac{r_h^2 \Omega^2}{4 c_s^2}, -\frac{r_h^2 \Omega^2}{4 c_s^2}, 1 - i \frac{r_h \omega}{c_s}, 0, 0; 1 - \frac{r^2}{r_h^2}\right),
\end{equation}
where we take a finite $r > r_h$ and the full solution is given by
\begin{equation*}
    \eta(t,r,\phi) = \sum_{\ell = 0}^\infty b_\ell \ \eta_\ell(t,r,\phi),
\end{equation*}
for constant coeficients $b_\ell$. For $r \geq 2r_h$, one should replace the power series representation for this local HeunC function by some analytic continuation of it \cite{Bartolomeu_2002,Bartolomeu_2005,Bartolomeu_2008,Bartolomeu_2015,Ishkhanyan_2005_Beta,slavyanov}. If $r \gg r_h$ or $r\to\infty$, one should instead use the asymptotic solution \eqref{solucao assintotica boa}.
%
\section{Results and discussion}\label{sec:resultados}
In this section, we apply the RFM to our analogue black hole problem. To do this, we need to substitute $q,\alpha', \gamma', \delta'$ and $\epsilon'$, as defined by equations \eqref{def Omega maiusculo} and by \eqref{parametros heun}, into the limit \eqref{limite heunc} to calculate the eigenvalues. 
%
%

We evaluate
%
\begin{equation}
    \left|\text{HeunC}\left(\frac{\ell^2}{4} -\frac{r_h^2 \Omega^2}{4 c_s^2}, -\frac{r_h^2 \Omega^2}{4 c_s^2}, 1 - i \frac{r_h \omega}{c_s}, 0, 0; 1 - u_r\exp\left[i\left(\pi - \phi\right)\right]\right)\right| = 0
\end{equation}
where $u = u_r e^{i\theta}$, $\alpha' = \alpha'_r e^{i\phi}$, and $u_r \gg 1$ for a given $\ell$. Additionally, we find that defining $\omega = \Re\omega - i \Im\omega$ facilitates the numerical evaluation of the equation. We solve this equation numerically using the HeunC function as implemented in Mathematica v. 13.1 and present the first five QNM frequencies below:
%
%
%
\begin{table}[H]
\centering
\begin{tabular}{|c|c|}
\hline
$\ell = 0$                & $\ell = 1$                \\ \hline
$0.170942848- i1.013561752$ & $0.303980463- i0.197735412$ \\ \hline
$0.191529062- i1.012599493$ & $0.371711061- i1.035347791$ \\ \hline
$0.238445 - i 1.00959     $ & $0.740535356- i0.934393665$ \\ \hline
$0.239664854- i1.009442840$ & $0.745119309- i0.934074799$ \\ \hline
$0.278516454- i1.006133668$ & $0.756311932- i0.933214077$ \\ \hline
\end{tabular}
\caption{First five QNM frequencies using $u_r = 23$, $r_h = 0.9$, $c_s = 1$.}
\end{table}
The QNM frequencies $\omega_{\ell,m}$ (the $m$-th frequency of the $\ell$ mode) on this table are the ones that obey the boundary conditions and simultaneously solve the confluent Heun equation around the singularities located at the horizon and at infinity. Morever, they make it such that the solutions are indeed proportional to each other in the overlapping domain of both representations in power series, as in equations \eqref{frobenius} and \eqref{subnormal}. These frequencies may be interpreted as the characteristic frequencies in which the perturbations on the background flow are damped, with respect to time and radial distance to the origin. Thus, a background velocity field imposed by equation \eqref{eq euler ordem 0 n constante sistema 3} may be perturbed by a second external field such that its perturbations \textit{ringdown} exhibiting those characteristic frequencies, much like the ringdown of the gravitational waves produced by a black hole merger \cite{ringdown1,ringdown2,ringdown3}. The consequences of such perturbation are then described by the Klein--Gordon equations \eqref{equacao de klein-gordon plasmons superficial} or \eqref{equacao de klein-gordon plasmons superficial 2}, as well as equation \eqref{eq campo eletrico fast}. This last equation is of importance, because it means that every perturbation $n_1$, defined by equation \eqref{equacao de klein-gordon plasmons superficial}, produces an electric field as described by the perturbed electric field $\vec{\nabla} \cdot \mathbf{E}_1 = -e n_1/\varepsilon$ and therefore, may be of experimental interest. 
\section{Final remarks}\label{sec:conclusao}
This work explored the consequences of a thin--layer procedure to the fluid equations of a toy model electron gas, inspired by the works of da Costa \cite{dacosta}, Ferrari and Cuoghi \cite{Ferrari_2008}, and subsequential investigations \cite{kerr-de-sitter,dini,kerr-sen, valle}. We found that there exists a geometric potential acting on the surface perturbations of this gas, which can be applied to a finely tuned gapless scalar plasmon transport as well as to the Analogue Gravity programme. The results we find are a direct parallel to the conclusions drew by Sch\"utzhold and Unruh \cite{unruh-shallow}, but here applied to a semiclassical electron gas. In the analogue black hole example, we found that the perturbations may decay with quasinormal frequencies, imitating the famous black hole ringdown modes. Because of the electric nature of this system, we expect the perturbations in a real electron gas to produce an electric field which may be sensed in a laboratory. Such experiment could help us understand the consequences of this geometric potential while exploring yet another possibility for analogue gravity investigations. We believe that our model for this geometric potential, being simpler and less artificial, is more adequate then da Costa's, because of the arbitrarity of his confining potential, which may not even be measurable. Our results follow from a simple physical consideration encapsulated in a Robin boundary condition, which results in a different geometric potential and does not require a contrived potential to locate the perturbations on an interface. There are evidences supporting the existence of such geometry--induced potentials \cite{valle,bonsel}, which contributes to the analysis of this model.
\begin{acknowledgments}
M. E. P. would like to give a special thanks to A. G. M. Schmidt for the years of working together. 
\end{acknowledgments}
\appendix
\section{Integrating across the interface} \label{apendice:integral}
To facilitate the calculations, we refer to the variable $q_3$ as $q$ and define the factor 
\begin{equation}
    D(q) = 1 - 2H q + K q^2,
\end{equation}
so that the normal contribution of the Laplacian reads
\begin{equation}
    \mathcal{L}_3 = \partial_3^2 + A(q)\partial_3,
\qquad
A(q) = \frac{D'(q)}{D(q)}
       = \frac{-2H + 2K q}{1 - 2H q + K q^2}.
\end{equation}
%
The piecewise density function (for the normal dependence) is proportional to
\begin{equation}
f(q) =
\begin{cases}
e^{-\alpha_2 q}, & q > 0,\\[4pt]
e^{\alpha_1 q}, & q < 0,
\end{cases}
\qquad
\alpha_2 = \alpha,\quad
\alpha_1 = -\alpha \frac{\epsilon_2}{\epsilon_1}.
\end{equation}
We integrate over a symmetric layer $[-L,L]$ with $L \gg 1/\alpha$ (so that the exponentials have decayed at the boundaries) and $(L \ll H^{-1}, K^{-1/2}$. 

We choose an intermediate cutoff $L$ such that $ 1/\alpha \ll L \ll H^{-1}$. 
Since the integrands are exponentially localized within $|q_3| \lesssim 1/\alpha $, 
extending the integration limits to $\pm\infty$ introduces errors that are 
exponentially small in $\alpha L$, and hence negligible. Within this same region, 
the expansion of the metric factor $\ln D(q_3) $ to second order is uniformly valid.

First, we define the layer integrals of the normal part, divided into the positive and negative branches:
\begin{equation}
    I_2 = \int_0^\infty \mathcal{L}_3 f_2 \, dq,
\qquad
I_1 = \int_{-\infty}^0 \mathcal{L}_3 f_1 \, dq.
\end{equation}
Their integrands are, in the positive branch, or $q>0$,
\begin{equation}
   \mathcal{L}_3 f_2 
= \alpha_2^2 f_2 + A(q)(-\alpha_2 f_2)
= e^{-\alpha_2 q} \bigl(\alpha_2^2 - \alpha_2 A(q)\bigr),
\end{equation}
and in the negative branch, or $q<0$,
\begin{equation}
    \mathcal{L}_3 f_1 
= \alpha_1^2 f_1 + A(q)(\alpha_1 f_1)
= e^{\alpha_1 q} \bigl(\alpha_1^2 + \alpha_1 A(q)\bigr).
\end{equation}
We now calculate the positive integral, resulting in
\begin{equation}
    I_2 = \int_0^\infty e^{-\alpha_2 q} \bigl(\alpha_2^2 - \alpha_2 A(q)\bigr) dq.
\end{equation}
Since $A(q) = D'(q)/D(q) = d(\ln D(q))/dq$, integration by parts gives
\begin{equation}
    \int_0^\infty e^{-\alpha_2 q} A(q)\,dq
= \alpha_2 \int_0^\infty e^{-\alpha_2 q} \ln D(q)\,dq,
\end{equation}
with the boundary term vanishing because \(\ln D(0)=0\). Therefore,
\begin{equation}
    I_2 = \alpha_2 - \alpha_2^2 \int_0^\infty e^{-\alpha_2 q} \ln D(q)\,dq.
\end{equation}
As for the integral of the negative branch, we proceed as before, but we introduce $q = -x$, with $x>0$. Then,
\begin{equation}
I_1 = \alpha_1 - \alpha_1^2 \int_0^\infty e^{-\alpha_1 x} \ln D(-x),dx.
\end{equation}
%
%
Since $q \lesssim 1/\alpha \ll H^{-1}$ and $q \ll |K|^{-1/2}$, we expand
\begin{equation}
\ln D(q) = \ln(1 - 2Hq + Kq^2)
\approx -2Hq + (K - 2H^2)q^2,
\end{equation}
and
\begin{equation}
\ln D(-x) = \ln(1 + 2Hx + Kx^2)
\approx 2Hx + (K - 2H^2)x^2.
\end{equation}
Evaluating the positive integral, we find
\begin{equation}
\int_0^\infty e^{-\alpha_2 q} \ln D(q),dq
= \int_0^\infty e^{-\alpha_2 q} \left[-2Hq + (K-2H^2)q^2\right] dq
= -\frac{2H}{\alpha_2^2} + \frac{2(K-2H^2)}{\alpha_2^3}.
\end{equation}
Therefore,
\begin{equation}
I_2 = \alpha_2 - \alpha_2^2 \left(-\frac{2H}{\alpha_2^2} + \frac{2(K-2H^2)}{\alpha_2^3}\right)
= \alpha_2 + 2H - \frac{2(K-2H^2)}{\alpha_2}.
\end{equation}
Evaluating the negative integral, we find
\begin{equation}
\int_0^\infty e^{-\alpha_1 x} \ln D(-x),dx
= \int_0^\infty e^{-\alpha_1 x} \left[2Hx + (K-2H^2)x^2\right] dx
= \frac{2H}{\alpha_1^2} + \frac{2(K-2H^2)}{\alpha_1^3}.
\end{equation}
Thus,
\begin{equation}
I_1 = \alpha_1 - \alpha_1^2 \left(\frac{2H}{\alpha_1^2} + \frac{2(K-2H^2)}{\alpha_1^3}\right)
= \alpha_1 - 2H - \frac{2(K-2H^2)}{\alpha_1}.
\end{equation}
Adding $I_2$ and $I_1$, the linear curvature terms cancel exactly:
\begin{equation}
I_{\text{norm}} = I_2 + I_1
= \alpha_2 + \alpha_1 - 2(K-2H^2)\left(\frac{1}{\alpha_2} + \frac{1}{\alpha_1}\right).
\end{equation}
The integral of the density itself over the layer is
\begin{equation}
I_f = \int_{-L}^{L} f(q),dq
= \int_0^\infty e^{-\alpha_2 q},dq + \int_{-\infty}^0 e^{\alpha_1 q},dq
= \frac{1}{\alpha_2} + \frac{1}{\alpha_1}.
\end{equation}
The effective potential is defined with the help of the ratio $I_\text{norm}/I_f$, which reduces to
\begin{equation} \label{geometric potential apendice}
    \frac{I_\text{norm}}{I_f} = -\alpha_1^2 \frac{\epsilon_1}{\epsilon_2} + 4H^2 - 2K .
\end{equation}
We define
\begin{equation}
    \boxed{
    V_\text{geom} = c_s^2 \left(4H^2 - 2K \right)
    }.
\end{equation}
%



\bibliography{apssamp}

\end{document}